# Rainbow Trapping in a Chirped Three-Dimensional Photonic Crystal

Zeki Hayran[1], Hamza Kurt[1], and Kestutis Staliunas[2,3]

[1]Nanophotonics Research Laboratory, Department of Electrical and Electronics Engineering, TOBB University of Economics and Technology, Ankara 06560, Turkey

[2]DONLL, Departament de Física, Universitat Politècnica de Catalunya (UPC), Edifici Gaia, Rambla Sant Nebridi 22, 08222 Terrassa, Spain

[3]Institució Catalana de Recerca i Estudis Avancats (ICREA), Passeig Lluís Companys 23, 08010 Barcelona, Spain

**Correspondence:** zekihayran@etu.edu.tr

Light localization and intensity enhancement in a woodpile layer-by-layer photonic crystal, whose interlayer distance along the propagation direction is gradually varied, has been theoretically predicted and experimentally demonstrated. The phenomenon is shown to be related to the progressive slowing down and stopping of the incoming wave, as a result of the gradual variation of the local dispersion. The light localization is chromatically resolved, since every frequency component is stopped and reflected back at different spatial positions. It has been further discussed that the peculiar relation between the stopping distance and the wave vector distribution can substantially increase the enhancement factor to more than two orders of magnitude. Compared to previously reported one- and two-dimensional photonic crystal configurations, the proposed scheme has the advantage of reducing the propagation losses by providing a three-dimensional photonic bandgap confinement in all directions. The slowing down and localization of waves inside photonic media can be exploited in many applications that requires enhanced interaction of light and matter.

## 1. Introduction

Temporary storage of light, by means of its slowing down which leads to increase of the intensity of slow light, has been the subject of intensive research, since many useful optical phenomena rely on strong interaction between light and matter[1]. Initial attempts to dramatically reduce the speed of light were based on quantum effects such as electromagnetic induced transparency[2] and optical coherent effects[3], where restrictions like the usage of ultracold gases or the narrow operational bandwidth limited their practical implementations. More recently the on-chip realizations of slow light schemes have become possible, such as coupled resonator optical waveguides[4,5], which can lead to low group velocities with negligible distortion. However a fundamental limit, limiting such structures is the so-called delay-bandwidth product, which imposes a trade-off between the factor of slowing-down and the bandwidth in which the slowing occurs[6]. One way to alleviate this limitation is to adiabatically tune (or chirp) the structure, meaning that one or several structural parameters are gradually varied along the direction of propagation. In particular, it has been shown that in a tapered metamaterial heterostructure, light can be slowed down and trapped at specific locations depending on its frequency using negative Goos–Hänchen shift[7]. This phenomenon, termed as the "Trapped Rainbow" effect, can also be considered as the spatial separation of the frequency components of the propagating wave, and has been demonstrated in dielectric[8] and plasmonic[9-18] gratings or waveguides, in one-[19] and two-dimensional[20-22] photonic crystals (2D PhCs), in a hyperbolic metamaterial waveguide[23], in sonic crystals[24], and in a waveguide under a tapered magnetic field[25]. While significant steps have been taken towards the slowing down and "trapping" of the light, one still needs to cope with extrinsic losses while implementing such schemes. It has been debated whether material absorption that is reminiscent of metallic structures can give large propagation losses and prevent the light stopping mechanism in metamaterials[26]. Plasmonic structures also suffer from the losses, which imply the need of gain media[11]. Although it is possible to overcome the material losses in metamaterials by employing absorption-free dielectric 2D PhCs in their negative refraction regime[27] instead of metallic components, it still remains unclear whether other extrinsic loss mechanisms such as out-of-plane radiation losses can be overcome in such structures. Yet some attempts to reduce the losses in 2D PhCs have been undertaken[28,29], the inevitable relation between the group velocity of the light and its interaction time with the medium still leads to considerable amount of scattering loss enhancement, as one approaches the band edge region (where most of the slow light behavior occurs)[27,30-33].

In contrast to their 2D counterparts, three-dimensional (3D) PhCs are capable of providing complete photonic bandgaps (PBGs) in all three directions. Among them, the layer-by-layer (or namely the woodpile) PhC has been intensively studied due to its flexible micro- or nanofabrication requirements and its robustness against fabrication deviations[34,35]. Such a structure consists of layers of parallel rods or logs, where the rods in each consecutive layer are rotated by 90°. Additionally, each layer with the same orientation of rods is shifted relative to each other by a half of the in-plane period ($a$), forming a 3D PhC of the symmetry of a face-centered tetragonal lattice. In this study, we propose a woodpile PhC with gradually varied longitudinal periods in the stacking direction, to gradually slow down and stop the incoming wave. We first present the local dispersion analysis of such a configuration which suggest the choice of structural parameters. We then evaluate the intensity enhancement factors and calculate the spatial positions where the wave stops and the maximal field enhancement occurs, by means of instantaneous and steady-state field distribution analysis. Finally we verify the principle experimentally in the microwave regime, by comparing the numerically calculated and experimentally measured field profiles and the intensity enhancement factors.

2. **Materials and Methods**

We consider a 3D woodpile PhC composed of cylindrical alumina (Al$_2$O$_3$) rods with a radii and height equal to $r = \frac{\sqrt{2}}{8}a$ and $h = 8.463a$, respectively, where $a$ is the in-plane period. The Al$_2$O$_3$ rods have a dielectric constant equal to $\varepsilon_{\text{rod}} = 9.8$ at microwave frequencies. The general schematics and the main operational principle of the proposed configuration are shown in Fig. 1(a). As mentioned above, the period in the stacking direction is continuously varied. We consider a slow (adiabatic) variation of the PhC period, which allows to consider local dispersion relation of it at every layer[24]. In result, the PBG of the local dispersion becomes smoothly varying along the propagation direction, as shown in Fig. 1(b). Therefore, a wave entering the structure will gradually slow down inside the PhC as its frequency will progressively approach the local PBG. Once the PBG is reached, the wave will then momentarily "stop", i.e. become temporarily localized, and will afterwards be reflected back. Since the local PBG is distinct for every spatial position, each frequency component will be localized at different depths, as shown representatively in the left inset of Fig. 1(a).

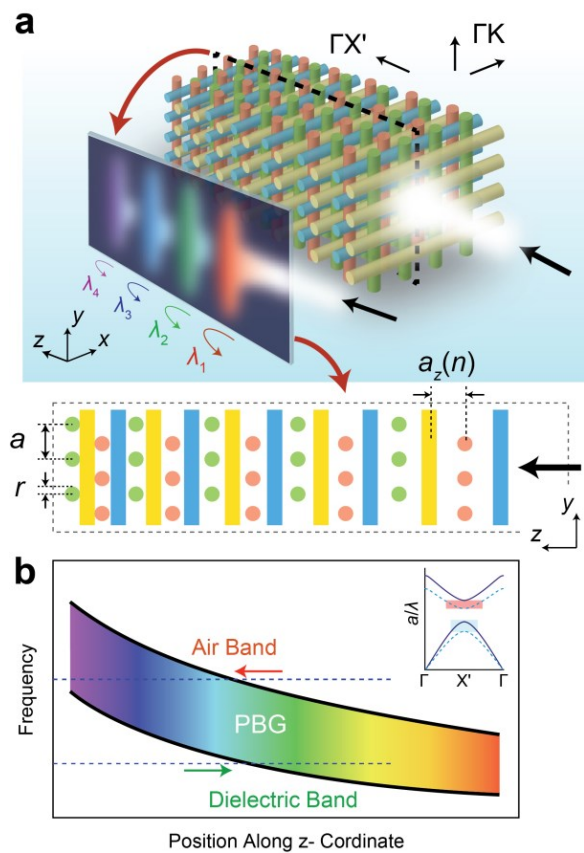

**Figure 1 (a)** Schematic illustration of the chirped woodpile PhC structure. For better visualization the number of periods of the illustrated PhC structure is reduced. The lower inset depicts the cross sectional view of the structure in the *yz*- plane. **(b)** The variation of the PBG along the propagation direction is representatively depicted. Due to the gradual variation of the layer-to-layer distances, the incident wave will be spatially separated into its frequency components along the propagation direction, thus forming a "trapped rainbow", as it is representatively shown as a cross sectional view on the lower left of **(a)**. The upper right inset in **(b)** representatively shows the change of the dispersion curves in the propagation direction, where the layer-to-layer distance of the solid black dispersion curve is smaller than that of the dashed blue dispersion curve. The black arrows reveal the propagation direction of the incoming wave.

A detailed numerical analysis is performed to deduce the structural parameters. Preliminary, for a periodical woodpile PhC with a constant layer-to-layer distance $d = \frac{\sqrt{2}}{4}a$ (*i.e.* the rods from the consecutive layers just touch one another) the local band structure is numerically calculated with the plane-wave expansion method[36] and the result is shown in Fig. 2(a). The calculated band structure reveals a complete PBG with a gap/mid-gap ratio (the ratio of gap width to the central frequency of the gap) of 11%, while the gap/mid-gap ratio of the partial PBG in the propagation direction Γ-X' (stacking direction) is equal to 39%. Moreover, the frequencies at the bottom and top of this partial PBG are defined as the lower and upper cut-off frequencies and their variation with respect to the layer-to-layer distance is plotted in Fig. 2(b). Despite their similarity in terms of their layer-to-layer distance relations, the two cut-off regions differ in an important respect: To obtain a local PBG at the lower and upper bands, the layer-to-layer distance should be increased and, conversely, decreased; respectively (see upper left inset of Fig. 1(b)). Consequently, the direction of the gradual layer-to-layer distance variation (whether it should be decreased or increased along the propagation direction) depends therefore on the operational frequency region: The upper and lower cut-off regions require a decreasing and increasing layer-to-layer distance variation, respectively (indicated by red and green arrows in Fig. 1(b), respectively). Apart from this apparent distinction, another difference between the two cut-off regions becomes evident if the partial PBGs in other directions are examined. The PBGs in the Γ-K direction, which corresponds to the propagation direction along the rod axes are superimposed on Fig. 2(b). Unlike the Γ-X' direction, the wave propagation in the Γ-K direction is nondegenerate for modes with polarization vectors along the rod axis and along the stacking direction, which we will refer to as the transverse electric-like (TE-like) and transverse magnetic-like (TM-like) modes, respectively. As can be seen from Fig. 2(b), the TM-like PBG comprises the whole upper cut-off region in the examined layer-to-layer distance range, whereas the TE-like PBG encloses only a portion of the examined range. On the other hand, for all utilized layer-to-layer distances, the lower cut-off region lies outside the coverage of both PBGs, which indicates that in the case of vertically scattered light, the corresponding structure based on the lower cut-off operation will be vulnerable to radiation losses. Furthermore, the PBG variations in the same range for propagation directions corresponding to the X, U, U', K', W, W', W", and L symmetry points[35], whose positions in the Brillouin zone are shown on the lower right inset of Fig. 2(a), are calculated and given in Fig. 2(c). For directions with nondegenerate modes, the complete PBGs comprising simultaneously both TE-like and TM-like PBGs have been taken into account. Consequently, the local dispersion analyses in Figs. 2(b) and 2(c) imply that a complete confinement is not possible, since modifying the layer-to-layer distances will further break the low rotational symmetry of the lattice structure[37], which in turn will increase the structural anisotropy. Nevertheless, a close inspection of the PBG variations will reveal that for the layer-to-layer distance interval $\{1.15, 1.35\} * \frac{\sqrt{2}}{4}a$, the localized mode near the cut-off region will still be confined in most directions.

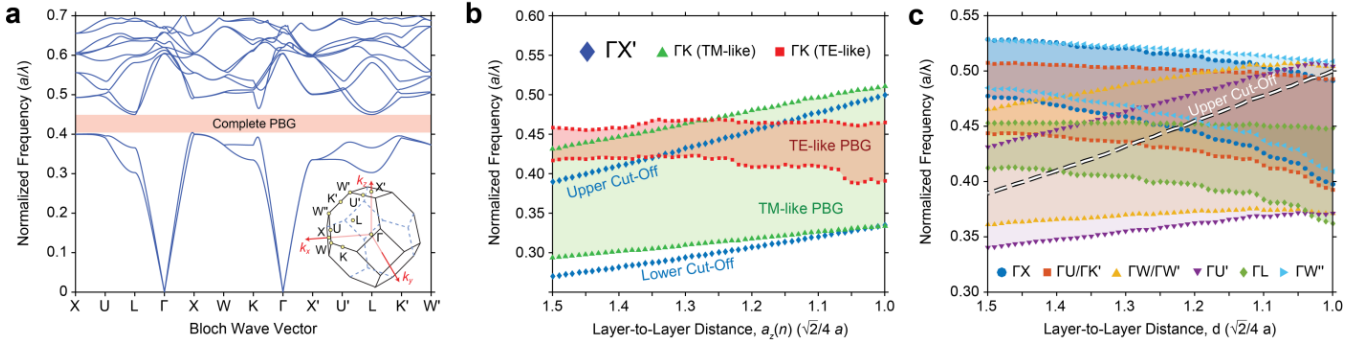

**Figure 2 (a)** Numerically calculated band structure of the woodpile PhC with $Al_2O_3$ rods. The lower right inset depicts the first Brillouin zone of the woodpile structure. **(b)** The dependency of the layer-to-layer distance onto the longitudinal and transverse cut-off frequencies. The TM-like and TE-like PBGs in the transverse directions are highlighted with light green and red colored areas. **(c)** The PBG variations for other directions. The dashed white line indicates the upper cut-off frequencies in the Γ- X' direction.

Following this approach, it seems reasonable to vary the layer-to-layer distances of the PhC structure from $1.35 * \frac{\sqrt{2}}{4} a$ to $1.15 * \frac{\sqrt{2}}{4} a$. However, one must take into account that the direct coupling into the slow light mode from the air region is problematic due to the mode mismatch at the air- PhC interface[38,39]. Furthermore, in order to avoid photon tunneling into the air cladding, one needs additional layers at the end of the structure. These two requirements can be both satisfied by increasing the layer-to-layer distance interval further including $\{1.00, 1.50\} * \frac{\sqrt{2}}{4} a$. In this way, the wave enters the structure with a wave vector away from the slow light region and also the leakage due to the large penetration depth of the slow light modes is suppressed. On the other hand, since the operational frequency region is at the upper cut-off region, the layer-to-layer distance should be decreased along the propagation direction, as was discussed previously. Moreover, the number of rods in each layer is chosen to be 7, as it is restricted by the length of the $Al_2O_3$ rods ($h = 8.463a$). The number of layers, on the other hand, is chosen to be 30, to create a smoothly varying chirp function. In this case, the local layer-to-layer distance between the $n$th and ($n$+1)th layer is equal to $a_z(n) = [(42.5 - 0.5n) * \frac{\sqrt{2}}{4} a]/28$.

### 3. Results and Discussion

To verify the propagation characteristics, the proposed structure was modeled in a 3D grid by employing commercially available software[40] based on the finite-difference time-domain (FDTD) method. Perfectly matched layer (PML) boundary conditions were used to terminate the computational domain and a TM polarized (electric field is in the *y*- direction) Gaussian pulse with a pulse length equal to *ct*/*a* = 30 in normalized time units was launched. The source was placed with a distance of 0.2*a* before the entrance of the structure. Figures 3(a), (b) and (c) show the electric field intensities, calculated along the propagation (*z*-) direction, at the center point of the *xy*- plane at each time frame for normalized central frequencies $a/\lambda = 0.450$, 0.468 and 0.496, respectively. Figures 3(a), (b) and (c) suggest that the PhC exhibits distinctive localization regions for each frequency. More precisely, the incoming wave is slowed down until it reaches its localization region (or turning point) and is then trapped for a finite time interval before it starts propagating backwards. If the trapping time is defined as the time duration of the electric field intensity decaying from its maximum value until its $1/e$ [12], then the normalized trapping times for Figs. 3(a), (b) and (c) would be obtained as $ct/a = 18.9$, 20.3 and 24.1; respectively. Further inferences can be made from the time evolution of the pulses, if one pays attention to the change of the magnitude of the field intensity. At the localization region the intensity increases, since the local energy density is expected to be inversely proportional to the group velocity[20,21,41]. Furthermore, to obtain a broadband analysis of the field enhancement, the steady-state electric field intensity was numerically calculated at the same region (along the *z*- propagation direction, at the center of the *xy*- plane), and is shown in Fig. 3(d). If the turning point is defined as the spatial position where the intensity becomes maximum, then a nearly linear increase of the position of the turning point is observed, which originates from both facts that the upper cut-off region is nearly linearly dependent on the layer-to-layer distance (see Fig. 1(b)) and that the chirp function is varying linearly. One interesting phenomenon which can be inferred from Fig. 3(d) is that the intensity enhancement is highly sensitive to the operational frequency. In particular, for the normalized operational frequency interval $a/\lambda = \{0.42, 0.50\}$ the maximum normalized intensities occurring at the localization regions deviate between 8 and 132. We attribute this deviation to the oscillations occurring between the turning point and the entrance of the structure. For instance, Figs. 3(a), 3(b) and 3(c) suggest that the intensity enhancement is nearly equal for these frequencies, which is expected since the nearly linear cut-off variation will induce similar slowdowns for different frequencies, even though their steady-state intensity enhancements are much diversified, as indicated with white arrows in Fig. 3(d). The main difference between these two cases is that the steady state analysis is the response of a continuous source excitement, which comprises multiple enhancement factors such as slowing down and multiple interferences of the reflected waves from the turning point and at the entrance of the structure; whereas the time evolution analysis is the response of a source with a finite temporal pulse width, which includes only the slowing mechanism in terms of the intensity enhancement. We therefore believe that the oscillatory behavior of the intensity enhancement stems from the peculiar relation between the optical path and the local wave vector distribution. In this regard, for a specific frequency the local wave vector distribution may lead to such a local wavelength distribution that it will encounter constructive interferences within the optical path (which is also unique to the operational frequency) and will give rise to an additional enhancement factor. Such frequencies correspond to the narrowband peaks in Fig. 3(d) and one can infer from this figure that the distance between these peaks decreases as the frequency increases. This is also reasonable, since increasing the frequency will cause the wave to localize and reflect further away from the entrance of the structure and to enter the structure with a wave vector away from the cut-off region, compared to lower operational frequencies. This will have the consequence that the wavelength distribution will initiate with lower spatial wavelengths as the wave enters the structure[21], in which case it will be enough to increase the operational frequency lesser compared to lower frequencies, to match up the next constructively interfered frequency.

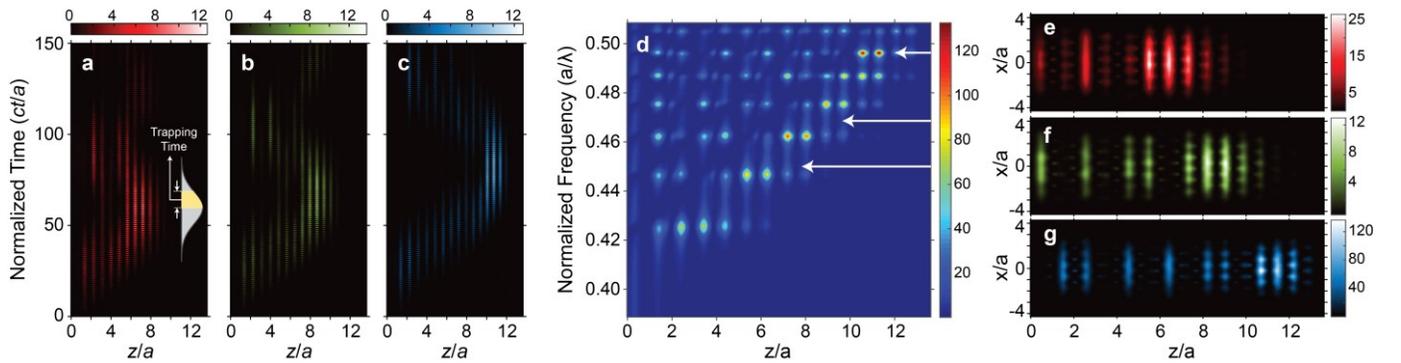

**Figure 3** Time evolution of a Gaussian pulse propagating inside the 3D PhC (along the *z*- propagation direction, at the center of the *xy*-plane), for three frequencies: **(a)** $a/\lambda = 0.450$, **(b)** $a/\lambda = 0.468$ and **(c)** $a/\lambda = 0.496$. **(d)** Broadband steady-state field intensity distributions at the same propagation line. Steady-state field distributions obtained at the *xz*- plane of the center of the *y*- direction for the normalized operational frequencies equal to **(e)** $a/\lambda = 0.450$, **(f)** $a/\lambda = 0.468$ and **(g)** $a/\lambda = 0.496$. The magnitude of the field intensities are normalized by dividing the calculated intensities to the intensity produced by the source at the specific frequencies. The distances in the *z*- propagation direction are given with respect to the position of the source.

Furthermore, the steady state electric field intensities along the *xz*- plane at the middle of the structural length in the *y*- transverse direction was calculated and are shown in Figs. 3(e), 3(f) and 3(g) for the normalized operational frequencies equal to $a/\lambda =$ 0.450, 0.468 and 0.496; respectively. One can see from these figures that the intensity enhancements differentiate that from the pulse excitations, as expected. Furthermore, additional localization regions can be observed apart from the localization region at the turning point, since the abovementioned resonator effect will give rise to a standing wave within the respective optical path. Nevertheless, the local energy density at the turning points will be higher than at the additional localization regions, as the slowing down will become maximum at these points[21].

To quantitatively evidence our interpretation regarding the additional enhancement mechanism described above, we calculated the phase of the different frequency components at a particular position. Figure 4 shows the unwrapped phase of the electric field in the *y*- direction at the longitudinal position equal to $10.55a$ with respect to the front face of the structure. Moreover, the normalized intensity spectrum at the same spatial position is superimposed on Fig. 4. One can deduce from this figure that for the frequencies, at which the intensities are locally maximum, the phases are integer multiples of $\pi$. This is in accord with our abovementioned interpretation of the large enhancement factors, since a constructive interference between a specific path requires also the same amount of phase shift[42].

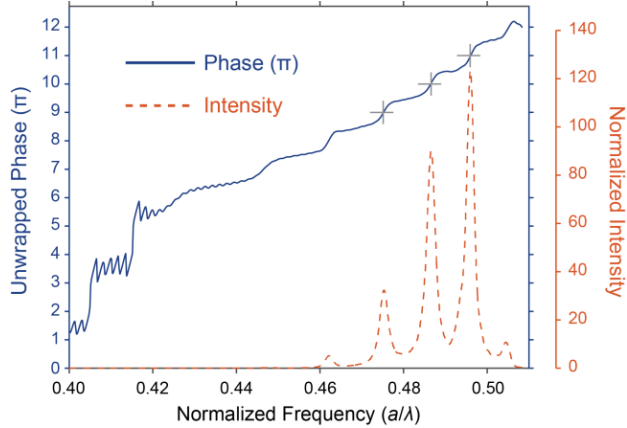

**Figure 4** Superimposed phase and normalized intensity spectrum. The unwrapped phases are calculated with respect to the phase at the entrance of the structure. The gray cross markers indicate the locations of the intensity peaks.

Further numerical analyses were carried out to reveal the quantitative amount of propagation losses which occur due to the finite transverse dimensions of the structure. The propagation losses were characterized by calculating the reflected power, by taking into consideration that in the case of a lossless propagation all incident power should be eventually reflected back after it is localized. The calculated reflected power is then normalized by the incident source power to give quantitative information about the loss spectra. Strictly speaking, since every frequency component will travel a different distance, it is much more physically meaningful to normalize the loss to the distance travelled by the wave for every frequency component. The travelling distance has been defined as the twofold of the distance between the entrance of the structure and the turning point, by taking into account the forward and backward propagation distances. In reality the travelling distance can be larger than the defined one, since the wave can have multiple roundtrips inside the structure. Therefore, for a more accurate analysis one must take this fact into account. However, in the present calculations we ignore this fact and define the travelling distance as an effective distance travelled by the wave, excluding the additional round trips. Following this direction, the loss spectra is calculated for 7 (original structure) and 12 (extended structure) transverse periods and are superimposed in Fig. 5. It can be observed from this figure that for the normalized frequency interval $a/\lambda = \{0.44, 0.48\}$ the propagation losses are lower than the rest of the operational frequency spectrum, which is expected as the TE-like transverse PBG does not compromise outside this interval, as was discussed. Another conclusion that can be drawn from Fig. 5 is that increasing the number of the transverse periods reduces the losses, as expected. It is worth noticing that the narrowband enhancement peaks in Fig. 3(d) matches well with the loss peaks in Fig. 5. As it was discussed, the narrowband enhancement peaks arises from multiple roundtrips inside the structure, which lies also in the origin of the enhanced losses.

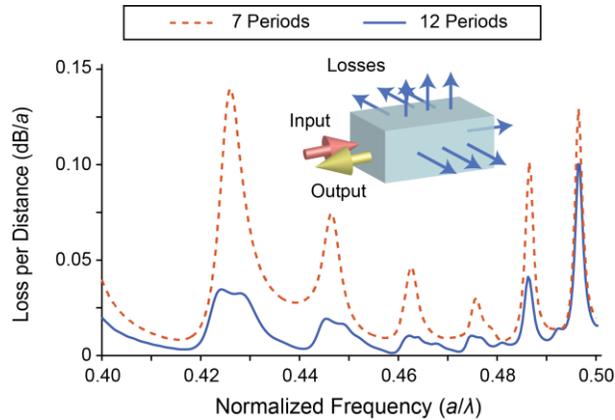

**Figure 5** Superimposed loss spectra of the PhC structure, with 7 and 12 transverse periods is shown. Note the oscillatory behavior, which is a direct consequence of the resonant propagation of the wave between the stopping point and the input side of the structure.

To verify the propagation characteristics experimentally in the microwave regime, the proposed structure was constructed with absorption free $Al_2O_3$ rods. Plates made of Plexiglas were employed to embody the constructed structure. The refractive index of Plexiglas has been measured via Bragg condition[43] to be equal to 1.59 and numerical analyses revealed that placing the Plexiglas plates does not change the results significantly. The structural parameters are same as in the numerical calculations, where the in-plane period is set equal to $a = 17.96$ mm. The experimental setup, which is shown in Fig. 6, consists of a standard horn antenna, a monopole coaxial antenna and a vector network analyzer (Agilent 5071C ENA). The horn antenna has an aperture size of 12.5 cm and 9.5 cm in the $x$- and $y$- directions, respectively, and is used to inject electromagnetic waves into the structure with electric field polarized in the $y$- direction, whereas the monopole antenna is used to detect the radiated electromagnetic waves. Furthermore, the experimental setup was covered with microwave absorbers to minimize the noise due to environmental reflections. In contrast to the calculations, in our experimental setup we can scan the electric field distribution only at the top surface of the structure. The monopole antenna was kept parallel to the $y$- axis and the tip of the monopole antenna was placed 2 mm above the top of structure. A motorized linear stage was used to step the monopole antenna with 1 mm lateral resolutions in the $x$- and $z$- directions, as shown in Fig. 6. In this way, the $y$- component of the electric field at the top surface of the structure was measured and shown in Figs. 7(a) and 7(b) for frequencies equal to 7.52 GHz ($a/\lambda = 0.450$) and 7.91 GHz ($a/\lambda = 0.473$), respectively. On the other hand, the electric field distribution was numerically calculated at the same surface and is given in Figs. 7(c) and 7(d) for frequencies equal to 7.52 GHz and 7.91 GHz, respectively. Comparing Figs. 7(a) with 7(c) and 7(b) with 7(d), we note that the turning points in the numerical and experimental cases are in close agreement. We attribute the discrepancy in the field distributions between the two cases to the difference of the experimental and numerical source beam shapes.

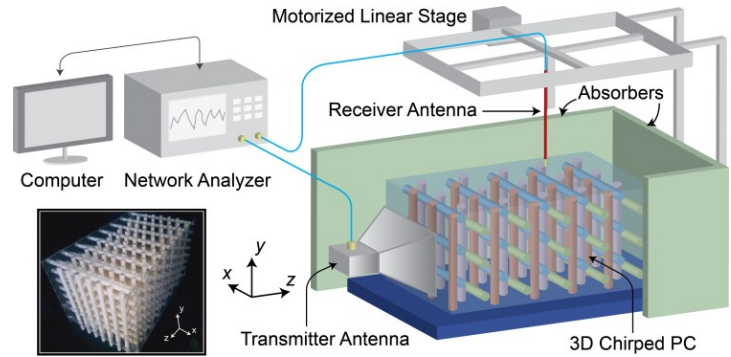

**Figure 6** Schematic illustration of the experimental setup is shown. The lower left inset depicts the photographic view of the constructed woodpile PC.

In view of the fact that the field distribution obtained at the top surface of the structure would not give any implication about the intensity enhancement, we further measured the electric field intensity spectrum at single points inside the structure to estimate the enhancement factors. For that purpose, the monopole antenna was inserted into the structure until the tip of the antenna reaches the center of the structural height in the *y*- direction. Keeping in mind that the monopole antenna should be parallel to the polarization axis of the injected electric field, the fact that the electric field is localized mostly between the rods that are perpendicular to the polarization axis inhibits the monopole antenna to be inserted parallel to the polarization axis. Therefore, as shown in Fig. 7(e) the monopole antenna was inserted from the neighboring layer inside and was tilted towards the region of interest with an angle of 16°. The measured electric field intensity was then normalized by the intensity measured with the same tilt angle at the center of the aperture of the transmitter horn antenna. Accordingly the normalized intensities were obtained at the center of the structural heights in the *x*- and *y*- directions and at a distance in the propagation *z*- direction equal to $5.15a = 9.25$ cm and $9.55a = 17.15$ cm with respect to the input side of the structure, and are superimposed with the regarding numerical calculations in Figs. 7(f) and 7(g), respectively. One can infer from these figures that the experimentally measured maximum intensities are around 72-74 for both cases, whereas the numerically calculated intensities are around 88-89. Moreover, one can observe that there are some spectral peak shifts and intensity differences between the numerical and experimental cases. We attributed these discrepancies to the structural fabrication errors, the non-uniform detection efficiency of the monopole antenna, the modified component of the detected electric field due to antenna tilting and the positional error of the detected area. Nevertheless, the experimental results verified that the wave can be enhanced and localized at various positions depending on its frequency.

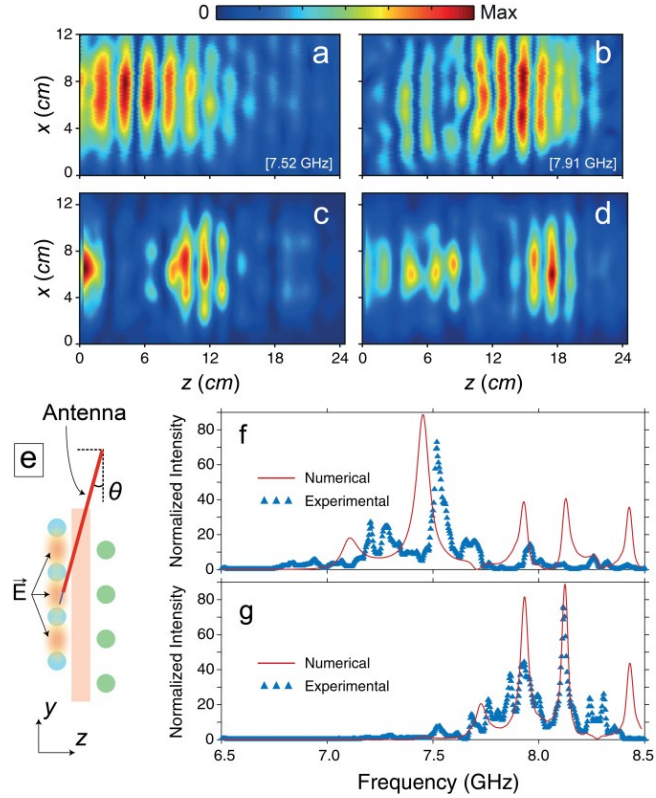

**Figure 7 (a-b)** Experimentally measured and **(c-d)** numerically calculated steady state electric field distributions at the top *xz*- surface of the structure for frequencies **(a-c)** 7.52 GHz and **(b-d)** 7.91 GHz are shown. **(e)** Schematic description of the measurement of the electric field intensity inside the structure. The numerically calculated and experimentally measured electric field intensity spectra, obtained at the center of the *xy*- plane and at a distance equal to **(f)** 9.25 cm and **(g)** 17.15 cm in the *z*- direction with respect to the input of the structure are superimposed.

4.  **Conclusions**

In summary, a chirped woodpile PhC, whose layer-to-layer distances are gradually decreased along the propagation direction, has been numerically and experimentally demonstrated to slow down and finally trap and enhance different frequency components at different spatial positions. It has been argued that the Fabry-Perot interferences occurring between the turning points and the entrance of the structure together with the adiabatic slowing mechanism can lead to intensity enhancements close to two orders of magnitude. An experimental realization at the microwave regime verified the operational principle, and we further note that the proposed structure is also feasible for current nanofabrication technologies, such as electron-beam lithography[44] or direct laser writing techniques[45,46], due to its simple layer-by-layer architecture and its robustness against fabrication disorders[34,35]. Furthermore we showed that the propagation losses can be suppressed, owing to the 3D periodicity of the structure. Such a phenomenon could be exploited to realize nonlinear optical devices, broadband photon harvesting systems, wavelength division multiplexing devices and optical buffers. We should note however, that the light phenomenon in the proposed configuration should not be considered as a permanent storage of a 'trapped rainbow'[7], but rather a temporary localization, due to the finite trapping time caused by the reflection at the turning points[25,47].

**Acknowledgements**

Authors acknowledge financial support of NATO SPS research grant No: 985048. H.K. also acknowledges partial support of the Turkish Academy of Science.